# $\eta$ Production in Peripheral Heavy-Ion Collisions

Alec J. Schramm *

and

Daniel H. Reeves

*Department of Physics, Occidental College, Los Angeles, CA 90041*

## Abstract

We estimate the impact parameter dependence of the production cross section for $\eta_c$ and $\eta_b$ mesons in peripheral heavy-ion collisions collisions. Total and elastic $\gamma\gamma$ cross sections are calculated in an equivalent photon approximation.

13.75.Cs,25.75.+r

Typeset using REVTEX

---

*Email: *alec@phys.oxy.edu*

Central collisions of heavy ions in facilities such as the Relativistic Heavy Ion Collider (RHIC) at Brookhaven and the Large Hadron Collider (LHC) at CERN will be crucial in the search for a quark-gluon plasma. With center-of-mass energies as high as 100 GeV/nucleon at RHIC and 3.5 TeV/nucleon at the LHC, these machines will have the additional feature of endowing the colliding nuclei with large virtual photon luminosities. This raises the prospect of studying electromagnetic particle production via two photon interactions [1]. Indeed, this mechanism has been used by several authors to calculate the cross section for $\gamma\gamma$ fusion to an intermediate-mass Higgs boson [2,3]. Another interesting possibility is the production of $c\bar{c}$ and $b\bar{b}$ meson states. As with the Higgs, the detection of such processes requires the low background attendant in peripheral ion collisions. In this paper, we calculate the production cross sections for the pseudoscalar mesons $\eta_c$ and $\eta_b$, and complement earlier work by Natale [4] with a calculation of their impact parameter dependence and the effects of nuclear absorption.

Our calculation is based upon the equivalent photon approximation (EPA) [5], in which one replaces the electromagnetic field of a relativistic charged particle with an equivalent pulse of real photons. For ion beams $A$ and $B$ with center of mass energy $s$, one assumes that the cross section for $\gamma\gamma$ fusion can be written in the form (see Figure 1)

$$\sigma_{AB}^{\gamma\gamma} = \int dx_1 dx_2 \, f_\pi^A(x_1) f_\pi^B(x_2) \sigma_{\gamma\gamma}^X(s_{\gamma\gamma}) \ . \tag{1}$$

In this expression $x_i$ is the fractional beam momentum carried by the virtual photons, and $\sigma_{\gamma\gamma}^X$ is the production cross section in the collision of two real photons with squared center of mass energy $s_{\gamma\gamma} \equiv x_1 x_2 s$ [6],

$$\sigma_{\gamma\gamma}^X = \frac{8\pi^2}{M_X} \Gamma_{X^0 \to \gamma\gamma} \, \delta(s - M_X^2) \ , \tag{2}$$

where $\Gamma_{X^0 \to \gamma\gamma}$ is the partial two-photon decay width of $X^0$. The function $f_\pi$ is the virtual photon distribution in the Coulomb field of a fast moving nucleus of charge $Z$, mass $M$, and nucleon number $A$ [3],

$$f_\gamma^A(x) = \frac{(Ze)^2}{\pi x} \int_0^\infty \frac{d^2 k_\perp}{(2\pi)^2} \, k_\perp^2 \, \frac{F_A(-k^2)^2}{(-k)^2} \tag{3}$$

$$= \frac{Z^2\alpha}{\pi x} \int_0^\infty d(k_\perp^2)\ k_\perp^2\ \frac{F_A(x^2M^2+k_\perp^2)^2}{(x^2M^2+k_\perp^2)^2}\ , \tag{4}$$

where $F_A(k^2)$ is the elastic nuclear form factor. In the language of the parton model, $f_\gamma(x)$ is the elastic photon distribution of the nucleus. It is the factor of $Z^2$ in $f$—which is compounded to $Z^4$ in the cross section—which renders electromagnetic interactions of high-$Z$ ions an effective tool for the production of heavy neutral particles.

Unlike earlier work on Higgs production, the present calculation must account for the pseudoscalar nature of $\eta$ mesons. In particular, if we denote by $\Gamma_{\mu\nu}$ the effective $\gamma-\eta$ vertex and in the high-energy limit keep only the transverse polarizations of the virtual photons, the coupling of an $\eta$ to the coherent nuclear electromagnetic current can be expressed in the form

$$(p_1+p_1')^\mu \Gamma_{\mu\nu}(p_2+p_2')^\nu \approx 4\sqrt{2}\frac{|\mathbf{k}_{1\perp}\times\mathbf{k}_{2\perp}|}{x_1 x_2}\overline{M}_{\gamma\gamma}(k_1,k_2)\ , \tag{5}$$

where $\overline{M}_{\gamma\gamma}$ is the spin-averaged invariant matrix element for the process $\gamma\gamma \to \eta$. The factor $|\mathbf{k}_{1\perp}\times\mathbf{k}_{2\perp}|$ is an expression of the pseudoscalar nature of the $\eta$ mesons; it would be replaced with $\mathbf{k}_1\cdot\mathbf{k}_2$ for a scalar final state such as a Higgs. In the spirit of the eikonal approximation, we then write the total $\gamma\gamma$ exchange cross section (1) in the impact parameter representation and fold it with the probability that no inelastic interaction takes place other than double photon exchange. The dependence of the cross section on impact parameter $b$ is then found by integrating the squared matrix element over all spacetime coordinates save for the transverse distance $b$ between the nuclei. We consider $^{208}$Pb nuclei, and approximate the form factor $F(\vec{k}^2)$ by the Gaussian [7]

$$F(\vec{k}^2) = e^{-\vec{k}^2/Q_0^2}\ , \tag{6}$$

where $Q_0 \approx 60\sqrt{2}$ MeV. Using this form factor, we find the following analytic expression for the cross section:

$$\frac{d\sigma_{AB}^{\gamma\gamma\to\eta}}{d^2b} = \frac{1}{\pi}\int \frac{dx_1\,dx_2}{x_1\,x_2}\ \sigma_{\gamma\gamma}^\eta(x_1 x_2 s)\int\ dq\ q\ J_0(qb)\ e^{-q^2/Q_0^2}$$

$$\times \left[ A(x_1)B(x_2) + A(x_2)B(x_1) - \frac{q^2}{4Q_0^2} \left[ C(x_1)B(x_2) + C(x_2)B(x_1) \right] \right] \quad (7)$$

where

$$A(x) = e^{-2x^2 M^2/Q_0^2} \int_0^\infty d\xi \, \xi \, \frac{e^{-2\xi}}{b} \left[ \sqrt{\frac{a}{a-b}} - 1 \right] \quad (8)$$

$$B(x) = e^{-2x^2 M^2/Q_0^2} \int_0^\infty d\xi \, \xi \, \frac{e^{-2\xi}}{b} \left[ 1 - \sqrt{\frac{a-b}{a}} \right] \quad (9)$$

$$C(x) = e^{-2x^2 M^2/Q_0^2} \int_0^\infty d\xi \, \frac{e^{-2\xi}}{\sqrt{a(a-b)}} \,, \quad (10)$$

and

$$a \equiv \left( \xi + \frac{q^2}{4Q_0^2} + \frac{x^2 M^2}{Q_0^2} \right)^2 , \quad b \equiv \xi q^2/Q_0^2 \,. \quad (11)$$

We have calculated the total production cross section $\sigma_{AA}^{\gamma\gamma \to \eta}$ and the differential cross section $d\sigma_{AA}^{\gamma\gamma \to \eta}/d^2b$ for both $\eta_c$ and $\eta_b$ production in the collision of $^{208}$Pb nuclei at both LHC and RHIC energies. Taking $\Gamma_{\eta_c \to \gamma\gamma} = 6.3$ keV and $\Gamma_{\eta_b \to \gamma\gamma} = 0.41$ keV [8], our calculations yield a total cross section of $\sigma_{AA}(\gamma\gamma \to \eta_c) = 0.75$ mb and $\sigma_{AA}(\gamma\gamma \to \eta_b) = 0.64$ $\mu$b at LHC energy; at RHIC, the numbers are understandably lower: $\sigma_{AA}(\gamma\gamma \to \eta_c) = 8.6$ $\mu$b and $\sigma_{AA}(\gamma\gamma \to \eta_b) = 0.17$ nb, respectively. These numbers are somewhat larger than those obtained by Natale [4], presumably due to our choice of form factor.

These cross sections, of course, are overly optimistic in that they do not account for the effects of inelastic nuclear scattering. The majority of the inelastic events is expected to occur at small values of the impact parameter $b$; indeed, the elastic nature of the interaction is maintained only in those collisions in which the two nuclei pass by each other. Thus it is important to verify that a significant portion of the $\gamma\gamma$ cross section extends out to relatively large impact parameters. For comparison, we have included inelastic scattering effects in two different ways. One is by applying a geometric cutoff at a minimum impact parameter of $2R$, where $R$ is the nuclear radius ($R \approx 7.1$ fm for $^{208}$Pb). A more realistic approach accounts for inelastic scattering effects using the Glauber approximation, with an absorption factor [9]

$$\frac{d\sigma_{AB}^{\text{el}}}{d^2b} = \frac{d\sigma_{AB}}{d^2b} \exp\left[-AB\ T_{AB}(b)\ \sigma_0\right], \quad (12)$$

where $\sigma_0$ is the total nucleon-nucleon cross section. (We have used the formula of Amaldi [10] to extrapolate $\sigma_0$ to RHIC and LHC energies.) For simplicity, we use the gaussian form factor (6) in calculating the profile function [11]

$$T_{AA}(b) = \int \frac{d^2Q}{(2\pi)^2}\ F_A(Q^2)\ F_A(Q^2)\ e^{iQb} = \frac{Q_0^2}{4\pi}\ e^{-Q_0^2 b^2/4}. \quad (13)$$

The dependence of the differential cross sections on impact parameter is shown in Figures 2 and 3, where the dotted curves show the suppressing effects of absorption in the Glauber approximation. (For comparison, the dot-dashed curves in Figure 2 represent the cross section for a scalar $\gamma\gamma - \eta$ coupling [3].) Our numerical results for both absorption methods are presented in Table 1.

There are, of course, backgrounds to this process which need to be considered, such as strong double diffractive scattering, i.e., double Pomeron exchange. These competing contributions are currently under investigation.

### ACKNOWLEDGMENTS

One of us (AJS) thanks Professor B. Müller for useful discussions. This research was supported by an award from Research Corporation.

**FIGURES**

FIG. 1:   The fusion of two virtual photons from scattering nuclei.

FIG. 2:   The impact parameter dependence of the differential cross section for $\gamma\gamma \to \eta_c$ in the nuclear scattering (a) at the LHC and (b) at RHIC. The solid curves are without absorption; the dotted curves include absorption effects. The dot-dashed curves show the result for a scalar $\eta$.

FIG. 3:   The impact parameter dependence of the differential cross section for $\gamma\gamma \to \eta_b$ in the nuclear scattering a) at the LHC and b) at RHIC. The solid curve is without absorption; the dotted curve includes absorption effects.

# TABLES

Table 1: Cross sections for $\eta$-meson production in peripheral collisions of $^{208}$Pb nuclei at RHIC and LHC energies. $\sigma_{AA}^{tot}$ is the total cross section. For comparison, inelastic scattering effects have been accounted for in two ways: $\sigma_{AA}(b > 2R)$ is the remaining cross section after applying a cut on impact parameter, whereas $\sigma_{AA}^{el}$ uses the Glauber approximation (see equation (12)).

Table 1

|  |  | $\eta_c$ | $\eta_b$ |
|---|---|---|---|
| 3.5 TeV/u (LHC) | $\sigma_{AA}^{tot}$ | 0.75 mb | 0.64 $\mu$b |
|  | $\sigma_{AA}(b > 2R)$ | 0.69 mb | 0.57 $\mu$b |
|  | $\sigma_{AA}^{el}$ | 0.69 mb | 0.57 $\mu$b |
| 100 GeV/u (RHIC) | $\sigma_{AA}^{tot}$ | 8.6 $\mu$b | 0.17 nb |
|  | $\sigma_{AA}(b > 2R)$ | 4.3 $\mu$b | 21.4 pb |
|  | $\sigma_{AA}^{el}$ | 4.6 $\mu$b | 26.5 pb |

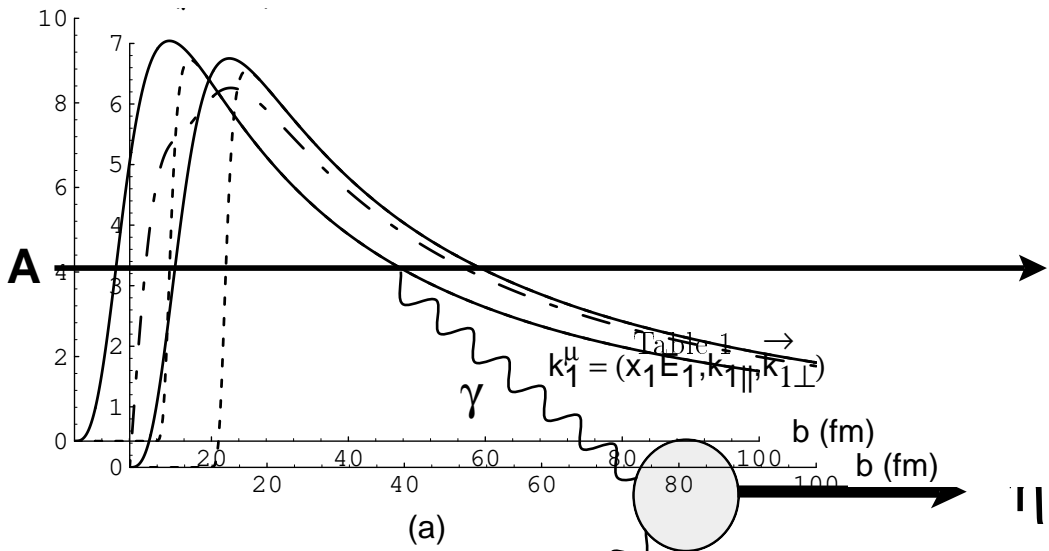

(a)

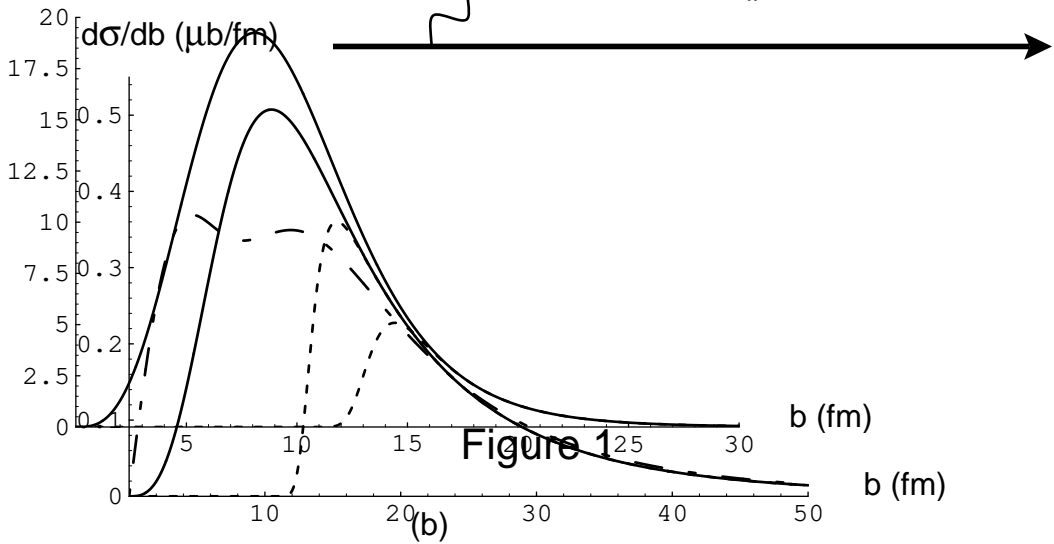

(b)

Figure 2

Figure 1